\documentclass[sigconf]{acmart}

\AtBeginDocument{%
  }

\usepackage{booktabs} 
\usepackage{bm}
\usepackage{multirow}
\usepackage{xspace}
\usepackage[tight,footnotesize]{subfigure}
\usepackage{algorithm}  
\usepackage{fancyhdr} 
\usepackage{amsmath}
\usepackage{enumitem}
\usepackage{algorithmic}
\usepackage{verbatimbox}

\newcommand{\ignore}[1]{}
\newcommand{\model}{LLM-KnowSimFuser}

\setcopyright{acmcopyright}
\copyrightyear{2024}
\acmYear{2024}
\acmDOI{XXXXXXX.XXXXXXX}

\acmConference[KDD'24]{Make sure to enter the correct
  conference title from your rights confirmation emai}{August,
  2024}{Barcelona, Spain}
\acmPrice{15.00}
\acmISBN{978-1-4503-XXXX-X/18/06}

\begin{document}

\title[Advancing Academic Knowledge Retrieval via LLM-enhanced Representation Similarity Fusion]{Advancing Academic Knowledge Retrieval via LLM-enhanced Representation Similarity Fusion: The 2nd Place of KDD Cup 2024 OAG-Challenge AQA}

\author{Wei Dai}
\affiliation{
\institution{Robo Space}
\country{}
}
\email{loveispdvd@gmail.com}

\author{Peng Fu}
\affiliation{
\institution{Robo Space}
\country{}
}
\email{fupeng@hotmail.com}

\author{Chunjing Gan}
\affiliation{
\institution{Ant Group}
\country{}
}
\email{cuibing.gcj@antgroup.com}


\begin{abstract}
In an era marked by robust technological growth and swift information renewal, furnishing researchers and the populace with top-tier, avant-garde academic insights spanning various domains has become an urgent necessity. 
The KDD Cup 2024 AQA Challenge is geared towards advancing retrieval models to identify pertinent academic terminologies from suitable papers for scientific inquiries. 
This paper introduces the {\model} proposed by {\itshape Robo Space}, which wins the 2nd place in the competition.
With inspirations drawed from the superior performance of LLMs on multiple tasks, after careful analysis of the provided datasets, we firstly perform fine-tuning and inference using LLM-enhanced pre-trained retrieval models to introduce the tremendous language understanding and open-domain knowledge of LLMs into this task, followed by a weighted fusion based on the similarity matrix derived from the inference results.
Finally, experiments conducted on the competition datasets show the superiority of our proposal, which achieved a score of 0.20726 on the final leaderboard.

\end{abstract}


\maketitle

\section{Introduction}
The overarching aim of scholarly data mining is to enhance our comprehension of the progression, essence, and direction of science. It possesses the capability to unveil substantial scientific, technological, and educational worth. 
In an age of vigorous technological expansion and rapid informational refreshment, equipping scholars and the general public with premier, cutting-edge academic knowledge across diverse disciplines is now an imperative demand.
The 2024 KDD Cup AQA competition is oriented toward enhancing retrieval algorithms with the aim to pinpoint relevant academic publications for scientific queries\cite{Zhang2024OAGBench}.
Inspired by the remarkable achievements of large language models (LLMs) such as ChatGPT \cite{chatgpt2023}, GPT4 \cite{gpt42023} in a variety of tasks due to their marvelous capability in language comprehension and generation,
in this paper, we introduces the {\model} solution proposed by {\itshape Robo Space}, which incorporates the tremendous language understanding and open-domain knowledge of LLMs in this solution and wins the 2nd place in the competition (achieved a score of 0.20726 on the final leaderboard) and organizes this technical report as follows:
\begin{itemize}[leftmargin=*]
\item First, we outline the task objectives and present the statistics of the given datasets in detail.
\item Subsequently, we introduce our data processing flow and process for fine-tuning and inference on LLM-enhanced pre-trained retrieval models with a carefully designed similarity fusion mechanism.
\item Finally, we conduct comprehensive ablation study and parameter analysis experiments on the competition datasets, which demonstrate the effectiveness and superiority of our proposal.
\end{itemize}

\section{Datasets}
The KDD Cup 2024 Academic Question Answering (AQA) Challenge is centered on tackling an academic retrieval problem. This endeavor employs a dataset that is systematically organized into two primary components: queries and documents. Queries embody academic questions, each structured with a concise title and an elaborative body that delineates the question's specifics. Documents, on the other hand, represent academic papers, each comprising a descriptive title and an informative abstract that encapsulates the paper's core contributions. 

All participants are required to navigate through two phases of the competition, with the latter building upon the former with enhanced complexity. The initial phase challenge contenders with a defined set of queries and documents, requiring the identification of the most pertinent documents for each query. Progressing to the second phase, the test set expands and the document collection significantly grows, escalating the task's intricacy while the core objective of discerning relevancy persists. We list the details of the statistics and objectives in Table \ref{tab:competition_summary}.

\begin{table*}[h]
\setlength{\abovecaptionskip}{0.25cm} \setlength{\belowcaptionskip}{0.25cm}
\setlength{\tabcolsep}{15pt}
\centering
\begin{tabular}{c|c|c|c|c}
\hline
Phase & Training Set & Test Set & Paper Collection & Objective \\
\cmidrule{1-5}
1 & 8,757 & 2,919 & 395,812 & Top 20 IDs per query \\
\cmidrule{1-5}
2 & Same as Phase 1 & 5919 (+3,000) & 466,387 (+70,575) & Same as Phase 1 \\
\hline
\end{tabular}
\caption{Competition phases statistics and objectives summary.}
\label{tab:competition_summary}
\end{table*}
\section{Methodology}

Our approach is composed of three main components:
\begin{itemize}[leftmargin=*]
    \item \textbf{Embedding Extraction using Pre-trained Models:} We employ several distinct LLM-enhanced pre-trained models to extract embeddings separately for queries and documents.
    \item \textbf{Tuning a Pre-trained Model:} Among the pre-trained models, one is fine-tuned. We then use this tuned model to extract embeddings for both queries and documents.
    \item \textbf{Similarity Matrix Computation and Fusion:} Embeddings from the above models (five in total, including the tuned one) are used to compute similarity matrices between queries and documents. We then fuse and rank these five similarity matrices to improve relevance assessment.
\end{itemize}

\subsection{Utilization of Pre-trained Retrieval Models}
\label{sec:pretrained}

We utilize four pre-trained models: NV-Embed-v1\cite{nvembed}, SFR-Embedding-Mistral\footnote{https://blog.salesforceairesearch.com/sfr-embedded-mistral/}, GritLM-7B\cite{gritlm-7b}, and Linq-Embed-Mistral\cite{LinqAIResearch2024}. All these models are based on the Mistral framework \cite{mistral}, which \textit{excel in capturing rich semantic relationships and contextual nuances with additional open-domain knowledge, leading to more accurate and relevant search results compared to traditional retrieval models}\footnote{During experiments, we incorporate traditional retrieval models such as BGE when testing, however the outcome is not promising which demonstrates the superiority of LLM-enhanced retrieval models in such specific domains.}. Besides, they share similar methods for prompt construction and embedding extraction.
Notably, the GritLM-7B model employs mean pooling by default, while the other three models utilize last token pooling. Although it is feasible to use different pooling techniques with GritLM, we adhere to mean pooling to remain consistent with the convention established during pre-training.

For document embedding extraction, embeddings can be directly obtained without additional prompt words. However, for queries, which are typically short and sparse in content, it is crucial to differentiate them from documents in a retrieval setting. Therefore, we experimented with various instructions and tags as prompts to enhance query embeddings, where we present the results of different configurations of tags and instructions in Section \ref{sec:exp}.

\subsection{Supervised Fine-Tuning of Retrieval Models}
\label{sec:finetune}

Among the evaluated models, the SFR-Embedding-Mistral model proved to be the most suitable candidate for fine-tuning due to its simplicity and inherent flexibility. We opt to use the Tevatron \cite{Tevatron} framework in conjunction with the Low-Rank Adaptation (LoRA) \cite{lora} method to optimize the model’s performance. For the fine-tuning process, we employed a comprehensive dataset comprising queries and academic papers in the training set, ensuring that the model can be well-adapted to handle the specific academic retrieval tasks with high accuracy.

To achieve this, we meticulously configure several key LoRA parameters. Specifically, we set the scaling factor to 64 and applied a dropout rate of 0.1 to prevent overfitting by randomly deactivating a fraction of neurons during training. Additionally, we define the rank of the low-rank matrices used for adaptation as 8. These configurations are chosen to strike a balance between model complexity and performance.
Notably, the model is fine-tuned for only one epoch. This decision is based on empirical evidence indicating that additional epochs of training led to a decline in performance, likely due to overfitting. By limiting the fine-tuning process to a single epoch, we are able to maintain the model's optimal performance and generalization capabilities without compromising its effectiveness.

\subsection{Similarity Fusion of Pretrained Models}

In this section, we describe the process of integrating similarity matrices derived from multiple models to achieve a unified and robust retrieval outcome. We utilize four pretrained models and one fine-tuned model, as detailed in Sections \ref{sec:pretrained} and \ref{sec:finetune}. The following steps will outline the detailed procedure for computing, normalizing, and fusing the similarity matrices.

Firstly, we compute the similarity matrices for the embeddings of queries and documents from each model independently. This involves measuring the similarity between the query embeddings and document embeddings generated by each model. We employ Faiss \cite{faiss}, an efficient library that leverages GPU acceleration, to expedite these similarity calculations. The use of GPU acceleration significantly enhances the computation speed, making it feasible to handle large-scale data efficiently.
Next, to ensure comparability across different models, we normalize the similarity matrices for each model on a per-query basis. Normalizing the similarity matrices ensures that the scores from different models are on a uniform scale, which is crucial for fair and effective fusion.
After normalization, we perform a weighted fusion of the similarity matrices. Each model’s normalized similarity matrix is assigned a weight that reflects its relative importance or performance, which combines the strengths of the individual models, leveraging their diverse perspectives to improve the robustness and accuracy of the similarity measurements.
Finally, based on the aggregated similarity scores, we rank the documents for each query. We identify and select the top 20 documents with the highest similarity scores as the final results for submission. This selection process ensures that the most relevant documents, as determined by the combined insights of multiple models, are presented as the output.

\section{Experiments}\label{sec:exp}

\subsection{Experimental Setup and Reproducibility}
All experiments were conducted on a machine equipped with an Intel(R) Xeon(R) Silver 4210 CPU @ 2.20GHz, 128GB of RAM, and an NVIDIA RTX A6000 GPU with 48GB of memory.
We fine-tune SFR-Embedding-Mistral model using LoRA on specific modules (q\_proj, k\_proj, v\_proj, o\_proj, down\_proj, up\_proj, and gate\_proj) with a learning rate of 1e-4, per-device batch size of 8, 1 epoch of training, query and passage lengths limited to 32 and 156 tokens, respectively.
To promote reproducibility, our source code is publicly available on GitHub\footnote{\url{https://github.com/loveisp/KDD\_2024\_AQA}}, providing comprehensive guidance on the operational processes. Detailed instructions on how to run the code can be found in the README.md file within the repository. 
Additionally, specific execution parameters and hyperparameters for each component are clearly outlined in their corresponding directories. 

\subsection{Performance Analysis of Individual and Fused Retrieval Models}

In this section, we analyze the impact of various retrieval models and the fused variant on performance across two evaluation phases. Table \ref{tab:model_scores} presents the best scores achieved in Phase 1 and Phase 2 by each model and our proposed {\model} which fuses the former models via similarity metrics, allowing for a comparative assessment of their effectiveness and robustness.

\begin{table}[h!]
\setlength{\abovecaptionskip}{0.2cm} \setlength{\belowcaptionskip}{0.2cm}
\centering
\begin{tabular}{c|c|c}
\hline
Retrieval Model & Phase 1 & Phase 2 \\
\cmidrule{1-3}
SFR-Embedding-Mistral & 0.20891 & 0.18659 \\
\cmidrule{2-3}
GritLM-7B & 0.20825 & 0.18622 \\
\cmidrule{2-3}
Linq-Embed-Mistral & 0.21208 & 0.18925 \\
\cmidrule{2-3}
NV-Embed-v1 & 0.21088 & 0.18315 \\
\cmidrule{2-3}
Fine-tuned SFR-Embedding-Mistral & 0.23160 & 0.17968 \\
\cmidrule{2-3}
{\model} & \textbf{0.24621} & \textbf{0.20726} \\
\hline
\end{tabular}
\caption{Performance comparison in two evaluation phases.}
\label{tab:model_scores}
\end{table}

The comparative analysis highlights that while different models exhibit varying degrees of effectiveness and stability, {\model} stands out with the highest scores in both evaluation phases. Its ability to maintain strong performance across diverse conditions makes it a highly effective and reliable retrieval model. This performance is indicative of the successful integration of LLM-enhanced representation similarity fusion, enabling more accurate and consistent academic knowledge retrieval. The inclusion of similarity fusion results further underscores the potential of combining multiple models to achieve superior performance, showcasing the benefits of an ensemble approach in enhancing retrieval tasks.

\subsection{Investigation into Configurations of Tags and Instructions}
In this study, we evaluate the performance of various model configurations by combining different tags and instructions. Table \ref{tab:configs} presents the results, showing how each combination impacted the retrieval performance in Phase 2 of this competition.
The detailed examination of Table \ref{tab:configs} leads to the following conclusions:

\begin{myverbbox}[\small]{\VerbContentSummaryD}
1. {title}\n{body}
2. <question_title> {title} </question_title>
   \n<question_body> {body} </question_body>
3. {title}. {body}
4. Title: {title}\nContent: {body}
5. <title> {title} </title>\n<content> {body} 
   </content>
\end{myverbbox}

\begin{table}[t]
	\centering
	\caption{Tag formatting details.}
	\label{tab:tags}
	\resizebox{\linewidth}{!}{\begin{tabular}{c}
	\toprule
        \VerbContentSummaryD       \\ \midrule
	\end{tabular}}
 \vspace{-0.5mm}
\end{table}

\begin{myverbbox}[\small]{\VerbContentSummaryD}
1. Given a question including title and body, 
   retrieve relevant papers that answer the 
   question.
2. Given a question including title and body, 
   retrieve the paper's title and abstract 
   that answer the question.
3. Given a web search query, retrieve relevant 
   passages that answer the query.
4. Given a question, retrieve passages that 
   answer the question.
\end{myverbbox}

\begin{table}[t]
	\centering
	\caption{Different instruction configurations.}
	\label{tab:instructions}
	\resizebox{\linewidth}{!}{\begin{tabular}{c}
	\toprule
        \VerbContentSummaryD       \\ \midrule
	\end{tabular}}
\end{table}

\begin{table}[t]
\setlength{\abovecaptionskip}{0.2cm} \setlength{\belowcaptionskip}{0.2cm}
\centering
\begin{tabular}{c|c|c|c}
\hline
Retrieval Model & Tag & Instruction & Score \\
\cmidrule{1-4}
SFR-Embedding-Mistral & 1 & 1 & 0.18390 \\
\cmidrule{2-4}
SFR-Embedding-Mistral & 1 & 2 & 0.18659 \\
\cmidrule{2-4}
SFR-Embedding-Mistral & 1 & 5 & 0.18503 \\
\cmidrule{2-4}
GritLM-7B & 2 & 1 & 0.18622 \\
\cmidrule{2-4}
GritLM-7B & 2 & 2 & 0.18367 \\
\cmidrule{2-4}
GritLM-7B & 2 & 4 & 0.18603 \\
\cmidrule{2-4}
Linq-Embed-Mistral & 4 & 1 & 0.18521 \\
\cmidrule{2-4}
Linq-Embed-Mistral & 4 & 2 & 0.18925 \\
\cmidrule{2-4}
Linq-Embed-Mistral & 4 & 3 & 0.18468 \\
\cmidrule{2-4}
Linq-Embed-Mistral & 4 & 4 & 0.18530 \\
\cmidrule{2-4}
NV-Embed-v1 & 1 & 1 & 0.18103 \\
\cmidrule{2-4}
NV-Embed-v1 & 3 & 1 & 0.18315 \\
\cmidrule{2-4}
NV-Embed-v1 & 4 & 1 & 0.18285 \\
\cmidrule{2-4}
NV-Embed-v1 & 4 & 2 & 0.18251 \\
\cmidrule{2-4}
NV-Embed-v1 & 4 & 3 & 0.18185 \\
\cmidrule{2-4}
NV-Embed-v1 & 4 & 4 & 0.18228 \\
\cmidrule{2-4}
NV-Embed-v1 & 4 & 5 & 0.18174 \\
\hline
\end{tabular}
\caption{Performance across diverse configurations of tags and instructions.}
\label{tab:configs}
\end{table}

\subsubsection{Tag Effectiveness}
\begin{itemize}[leftmargin=*]
\item \textbf{Tag 1 (${title}\textbackslash n{body}$)}: which demonstrates robust performance across multiple instructions, particularly with the SFR-Embedding-Mistral model, achieving the highest score of 0.18659 with Instruction 2. This suggests that this tag format is well-suited for models that process structured text effectively.
\item \textbf{Tag 2 ($<question\_title> {title}</question\_title>\textbackslash n<question\_body> {body} </question\_body>$)}: which exhibits strong results with the GritLM-7B model. The structured XML format appears to enhance the model's ability to parse and retrieve relevant information, as evidenced by the score of 0.18622 with Instruction 1.
\item \textbf{Tag 4 ($Title: {title}\textbackslash nContent: {body}$)}: which is the most versatile one, especially with the Linq-Embed-Mistral model. The highest overall performance score of 0.18925 was recorded with Tag 4 and Instruction 2, indicating that this tag format's clear separation of title and content is highly effective for this model.
\end{itemize}

\subsubsection{Instruction Impact}
\begin{itemize}[leftmargin=*]
\item \textbf{Instruction 2 (Given a question including title and body, retrieve the paper's title and abstract that answer the question.)}: which generally provides the best results across different tags and models. This instruction seems to align well with the models' retrieval mechanisms, suggesting that a focused retrieval objective (title and abstract) enhances performance.
\item \textbf{Instruction 1 (Given a question including title and body, retrieve relevant papers that answer the question.)}: which also performs well, particularly with Tag 1 and the SFR-Embedding-Mistral model. This indicates that a broader retrieval scope (entire papers) can be effective when paired with suitable tag formats.
\end{itemize}

\subsubsection{Model Specific Observations}
\begin{itemize}[leftmargin=*]
\item \textbf{SFR-Embedding-Mistral}: which consistently performs well with Tag 1 and different instructions, indicating its robustness and adaptability to this tag format.
\item \textbf{GritLM-7B}: which shows strong performance with Tag 2, highlighting its preference for well-structured tags. The model also performs well with Tag 4 and Instruction 4, suggesting a degree of flexibility in handling structured queries.
\item \textbf{Linq-Embed-Mistral}: which achieves the highest score overall, particularly with Tag 4 and Instruction 2. This combination's effectiveness underscores the importance of choosing the right tag-instruction pairing for maximizing model performance.
\item \textbf{NV-Embed-v1}: which shows consistent performance, however it does not achieve the highest scores compared to the other models. The highest score for NV-Embed-v1 was 0.18315 with Tag 3 and Instruction 1, indicating potential areas for optimization.
\end{itemize}

\subsubsection{Possible Directions for Future Work}
\begin{itemize}[leftmargin=*]
\item \textbf{Further Exploration of Untested Configurations}: there remains potential in exploring the full range of untested tag and instruction combinations. By systematically testing these configurations, it may be possible to discover even more effective pairings that are not covered in this study.
\item \textbf{Automated Prompt Generation}: developing automated systems to generate and test prompts dynamically could significantly enhance the efficiency of identifying optimal configurations. This approach would allow for a broader exploration of the parameter space and potentially uncover novel configurations that yield superior performance.
\item \textbf{Focused Optimization for Lower-Performing Models}: specific attention should be directed towards optimizing tags and instructions for models such as NV-Embed-v1. By understanding and addressing the limitations that led to lower performance, it may be possible to enhance the retrieval effectiveness of these models.
\item \textbf{Model-Specific Tailoring}: customizing tags and instructions based on the characteristics and strengths of individual models could further improve performance. For instance, models that excel with structured tags (like GritLM-7B) could benefit from even more refined tagging strategies.
\end{itemize}

In conclusion, the findings of this study underscore the significant impact of tag and instruction configurations on retrieval model performance, and strategic selection and optimization of these elements can lead to substantial improvements, and future research should continue to explore and refine these configurations to fully realize their potential.
\section{Conclusion}
In this paper, we presents our solution for the KDD Cup 2024 OAG-Challenge AQA. To tackle this task, we employ a multi-step approach. Firstly, we fine-tune and perform inference using LLM-enhanced pre-trained retrieval models, capitalizing on the powerful language understanding and retrieval capabilities of large language models. Next, we conduct a weighted fusion of the inference results, leveraging a similarity matrix derived from these results to optimize the retrieval performance. Through this meticulous process, our team, \textit{Robo Space}, achieves a commendable final score of 0.20726, and ranks the 2nd place on the final leaderboard.

\bibliographystyle{ACM-Reference-Format}
\bibliography{references}


\begin{thebibliography}{10}


\ifx \showCODEN    \undefined \def \showCODEN     #1{\unskip}     \fi
\ifx \showDOI      \undefined \def \showDOI       #1{#1}\fi
\ifx \showISBNx    \undefined \def \showISBNx     #1{\unskip}     \fi
\ifx \showISBNxiii \undefined \def \showISBNxiii  #1{\unskip}     \fi
\ifx \showISSN     \undefined \def \showISSN      #1{\unskip}     \fi
\ifx \showLCCN     \undefined \def \showLCCN      #1{\unskip}     \fi
\ifx \shownote     \undefined \def \shownote      #1{#1}          \fi
\ifx \showarticletitle \undefined \def \showarticletitle #1{#1}   \fi
\ifx \showURL      \undefined \def \showURL       {\relax}        \fi
\providecommand\bibfield[2]{#2}
\providecommand\bibinfo[2]{#2}
\providecommand\natexlab[1]{#1}
\providecommand\showeprint[2][]{arXiv:#2}

\bibitem[Gao et~al\mbox{.}(2022)]%
        {Tevatron}
\bibfield{author}{\bibinfo{person}{Luyu Gao}, \bibinfo{person}{Xueguang Ma}, \bibinfo{person}{Jimmy Lin}, {and} \bibinfo{person}{Jamie Callan}.} \bibinfo{year}{2022}\natexlab{}.
\newblock \showarticletitle{Tevatron: An Efficient and Flexible Toolkit for Dense Retrieval}.
\newblock \bibinfo{journal}{\emph{arXiv preprint arXiv:2203.05765}} (\bibinfo{year}{2022}).
\newblock


\bibitem[Hu et~al\mbox{.}(2022)]%
        {lora}
\bibfield{author}{\bibinfo{person}{Edward~J. Hu}, \bibinfo{person}{Yelong Shen}, \bibinfo{person}{Phillip Wallis}, \bibinfo{person}{Zeyuan Allen{-}Zhu}, \bibinfo{person}{Yuanzhi Li}, \bibinfo{person}{Shean Wang}, \bibinfo{person}{Lu Wang}, {and} \bibinfo{person}{Weizhu Chen}.} \bibinfo{year}{2022}\natexlab{}.
\newblock \showarticletitle{LoRA: Low-Rank Adaptation of Large Language Models}. In \bibinfo{booktitle}{\emph{ICLR}}.
\newblock


\bibitem[Jiang et~al\mbox{.}(2023)]%
        {mistral}
\bibfield{author}{\bibinfo{person}{Albert~Q. Jiang}, \bibinfo{person}{Alexandre Sablayrolles}, \bibinfo{person}{Arthur Mensch}, \bibinfo{person}{Chris Bamford}, \bibinfo{person}{Devendra~Singh Chaplot}, \bibinfo{person}{Diego de Las~Casas}, \bibinfo{person}{Florian Bressand}, \bibinfo{person}{Gianna Lengyel}, \bibinfo{person}{Guillaume Lample}, \bibinfo{person}{Lucile Saulnier}, \bibinfo{person}{L{\'{e}}lio~Renard Lavaud}, \bibinfo{person}{Marie{-}Anne Lachaux}, \bibinfo{person}{Pierre Stock}, \bibinfo{person}{Teven~Le Scao}, \bibinfo{person}{Thibaut Lavril}, \bibinfo{person}{Thomas Wang}, \bibinfo{person}{Timoth{\'{e}}e Lacroix}, {and} \bibinfo{person}{William~El Sayed}.} \bibinfo{year}{2023}\natexlab{}.
\newblock \showarticletitle{Mistral 7B}.
\newblock \bibinfo{journal}{\emph{arXiv preprint arXiv:2310.06825}} (\bibinfo{year}{2023}).
\newblock


\bibitem[Johnson et~al\mbox{.}(2021)]%
        {faiss}
\bibfield{author}{\bibinfo{person}{Jeff Johnson}, \bibinfo{person}{Matthijs Douze}, {and} \bibinfo{person}{Herv{\'{e}} J{\'{e}}gou}.} \bibinfo{year}{2021}\natexlab{}.
\newblock \showarticletitle{Billion-Scale Similarity Search with GPUs}.
\newblock \bibinfo{journal}{\emph{{IEEE} Trans. Big Data}} \bibinfo{volume}{7}, \bibinfo{number}{3} (\bibinfo{year}{2021}), \bibinfo{pages}{535--547}.
\newblock


\bibitem[Kim et~al\mbox{.}(2024)]%
        {LinqAIResearch2024}
\bibfield{author}{\bibinfo{person}{Junseong Kim}, \bibinfo{person}{Seolhwa Lee}, \bibinfo{person}{Jihoon Kwon}, \bibinfo{person}{Sangmo Gu}, \bibinfo{person}{Yejin Kim}, \bibinfo{person}{Minkyung Cho}, \bibinfo{person}{Jy yong Sohn}, {and} \bibinfo{person}{Chanyeol Choi}.} \bibinfo{year}{2024}\natexlab{}.
\newblock \showarticletitle{Linq-Embed-Mistral:Elevating Text Retrieval with Improved GPT Data Through Task-Specific Control and Quality Refinement}.
\newblock \bibinfo{journal}{\emph{Linq AI Research Blog}} (\bibinfo{year}{2024}).
\newblock


\bibitem[Lee et~al\mbox{.}(2024)]%
        {nvembed}
\bibfield{author}{\bibinfo{person}{Chankyu Lee}, \bibinfo{person}{Rajarshi Roy}, \bibinfo{person}{Mengyao Xu}, \bibinfo{person}{Jonathan Raiman}, \bibinfo{person}{Mohammad Shoeybi}, \bibinfo{person}{Bryan Catanzaro}, {and} \bibinfo{person}{Wei Ping}.} \bibinfo{year}{2024}\natexlab{}.
\newblock \showarticletitle{NV-Embed: Improved Techniques for Training LLMs as Generalist Embedding Models}.
\newblock \bibinfo{journal}{\emph{arXiv preprint arXiv:2405.17428}} (\bibinfo{year}{2024}).
\newblock


\bibitem[Muennighoff et~al\mbox{.}(2024)]%
        {gritlm-7b}
\bibfield{author}{\bibinfo{person}{Niklas Muennighoff}, \bibinfo{person}{Hongjin Su}, \bibinfo{person}{Liang Wang}, \bibinfo{person}{Nan Yang}, \bibinfo{person}{Furu Wei}, \bibinfo{person}{Tao Yu}, \bibinfo{person}{Amanpreet Singh}, {and} \bibinfo{person}{Douwe Kiela}.} \bibinfo{year}{2024}\natexlab{}.
\newblock \showarticletitle{Generative Representational Instruction Tuning}.
\newblock \bibinfo{journal}{\emph{arXiv preprint arXiv:2402.09906}} (\bibinfo{year}{2024}).
\newblock


\bibitem[OpenAI(2023a)]%
        {chatgpt2023}
\bibfield{author}{\bibinfo{person}{OpenAI}.} \bibinfo{year}{2023}\natexlab{a}.
\newblock \showarticletitle{Chatgpt: Optimizing language models for dialogue}.
\newblock


\bibitem[OpenAI(2023b)]%
        {gpt42023}
\bibfield{author}{\bibinfo{person}{OpenAI}.} \bibinfo{year}{2023}\natexlab{b}.
\newblock \showarticletitle{{GPT-4} Technical Report}.
\newblock \bibinfo{journal}{\emph{arXiv preprint arXiv:2303.08774}} (\bibinfo{year}{2023}).
\newblock


\bibitem[Zhang et~al\mbox{.}(2024)]%
        {Zhang2024OAGBench}
\bibfield{author}{\bibinfo{person}{Fanjin Zhang}, \bibinfo{person}{Shijie Shi}, \bibinfo{person}{Yifan Zhu}, \bibinfo{person}{Bo Chen}, \bibinfo{person}{Yukuo Cen}, \bibinfo{person}{Jifan Yu}, \bibinfo{person}{Yelin Chen}, \bibinfo{person}{Lulu Wang}, \bibinfo{person}{Qingfei Zhao}, \bibinfo{person}{Yuqing Cheng}, \bibinfo{person}{Tianyi Han}, \bibinfo{person}{Yuwei An}, \bibinfo{person}{Dan Zhang}, \bibinfo{person}{Weng~Lam Tam}, \bibinfo{person}{Kun Cao}, \bibinfo{person}{Yunhe Pang}, \bibinfo{person}{Xinyu Guan}, \bibinfo{person}{Huihui Yuan}, \bibinfo{person}{Jian Song}, \bibinfo{person}{Xiaoyan Li}, \bibinfo{person}{Yuxiao Dong}, {and} \bibinfo{person}{Jie Tang}.} \bibinfo{year}{2024}\natexlab{}.
\newblock \showarticletitle{OAG-Bench: A Human-Curated Benchmark for Academic Graph Mining}.
\newblock \bibinfo{journal}{\emph{arXiv preprint arXiv:2402.15810}} (\bibinfo{year}{2024}).
\newblock


\end{thebibliography}


\end{document}